\documentclass{article}
\usepackage{dcase2021,amsmath,graphicx,url,times,booktabs, tabularx}
\usepackage[table,xcdraw]{xcolor}

\def \am [#1]{\textcolor{red}{AM: #1}}
\def \wang [#1]{\textcolor{blue}{wang: #1}}

\title{Audio-visual scene classification:\\
analysis of DCASE 2021 Challenge submissions}
\name{Shanshan Wang, Toni Heittola, Annamaria Mesaros, Tuomas Virtanen\thanks{This work was supported in part by the European Research Council under the European Unions H2020 Framework Programme through ERC Grant Agreement 637422 EVERYSOUND.
}}
 \address{Computing Sciences \\ Tampere University, Finland\\
 \{shanshan.wang, toni.heittola, annamaria.mesaros, tuomas.virtanen\}@tuni.fi}
\begin{document}

\ninept
\maketitle

\begin{sloppy}

\begin{abstract}
This paper presents the details of the Audio-Visual Scene Classification task in the DCASE 2021 Challenge (Task 1 Subtask B). The task is concerned with classification using audio and video modalities, using a dataset of synchronized recordings. This task has attracted 43 submissions from 13 different teams around the world. Among all submissions, more than half of the submitted systems have better performance than the baseline. The common techniques among the top systems are the usage of large pretrained models such as ResNet or EfficientNet which are trained for the task-specific problem. Fine-tuning, transfer learning, and data augmentation techniques are also employed to boost the performance. More importantly, multi-modal methods using both audio and video are employed by all the top 5 teams. The best system among all achieved a logloss of 0.195 and accuracy of 93.8\%, compared to the baseline system with logloss of 0.662 and accuracy of 77.1\%.

\end{abstract}


\section{Introduction}
\label{sec:intro}

Acoustic scene classification (ASC) has been an important task in the DCASE Challenge throughout the years, attracting the largest number of participants in each edition. Each challenge included a supervised, closed set classification setup, with increasingly large training datasets \cite{Mesaros2016_EUSIPCO}, \cite{Mesaros2018_DCASE}, \cite{Mesaros2019}, which has allowed the development of a wide variety of methods. In recent years, the task has focused on robustness to different devices and low-complexity solutions \cite{heittola2020acoustic}.

Scene classification is commonly studied in both audio and video domains. For acoustic scene classification the input is typically a short audio recording, while for visual scene classification tasks the input can be an image or a short video clip. State-of-the-art solutions for ASC are based on spectral  features,  most  commonly the log-mel spectrogram, and convolutional neural network  architectures, often  used in large ensembles \cite{Mesaros2019}. In comparison, visual scene classification (VSC) from images has a longer history and more types of approaches, e.g. global attribute descriptors, learning spatial layout patterns, discriminative region detection, and more recently hybrid deep models \cite{Xie2020}. The classification performance for images has been significantly increased when large-scale image datasets like ImageNet \cite{imagenet_cvpr09} became available. Various network structures have been explored over these years, e.g. CNN \cite{krizhevsky2012imagenet}, while more recently, ResNet \cite{he2016deep} and EfficientNet \cite{tan2019efficientnet} have been proposed to further increase the performance. 

Motivated by the fact that we humans perceive the world through multiple senses (seeing and hearing), and in each individual domain methods have reached maturity, multimodal analysis has become a pursued research direction for further improvement.
Recent work has shown the joint learning of acoustic features and visual features could bring additional benefits in various tasks,
allowing novel target applications such as visualization of the sources of sound in videos \cite{Arandjelovic_2018_ECCV},  audio-visual alignment for lip-reading \cite{Chung2017}, or audio-visual source separation \cite{Zhao_2018_ECCV}. 
Feature learning from audio-visual correspondence (AVC) \cite{DBLP:journals/corr/ArandjelovicZ17}, and more recent work that learns features through audio-visual spatial alignment from 360 video and spatial audio \cite{morgado2020learning}, have shown significant improvement in performance  in various downstream tasks.
 
Audio-visual scene classification (AVSC) is introduced in DCASE 2021 Challenge for the first time, even though research on audio-visual joint analysis has been active already for many years. The novelty of the DCASE task is use of a carefully curated dataset of audio-visual scenes \cite{Wang2021_ICASSP}, in contrast to the use of audio-visual material from YouTube as in the other studies. Audio-visual data collected from the Youtube mostly has automatically generated label categories, which makes the data quality irregular. Besides, most of the datasets based on material from Youtube are task specific, e.g., action recognition \cite{soomro2012ucf101}, sport types \cite{Gade_2015_ICCV_Workshops}, or emotion \cite{zhang2018}. In \cite{Wang2021_ICASSP}, the dataset is carefully planned and recorded using the same equipment, which gives it a consistent quality. 

In this paper we introduce the audio-visual scene classification task setup of DCASE 2021 Challenge. We shortly present the dataset used in the task and the given baseline system. We then present the challenge participation statistics and analyze the submitted systems in terms of approaches. Since visual data has a large number of large datasets, e.g. ImageNet \cite{imagenet_cvpr09}, CIFAR \cite{krizhevsky2009learning}, most methods employing the visual modality are expected to use pretrained models or transfer learning based on the pretrained models. A number of resources have been listed and allowed as external data in the DCASE 2021 website.

The paper is organized as follows: Section 2 introduces the dataset, system setup and baseline system results. Section 3 presents the challenge results and Section 4 gives an analysis of selected submissions. Finally, Section 5 concludes this paper.

\begin{figure}
    \centering
    \includegraphics[width=0.9\columnwidth]{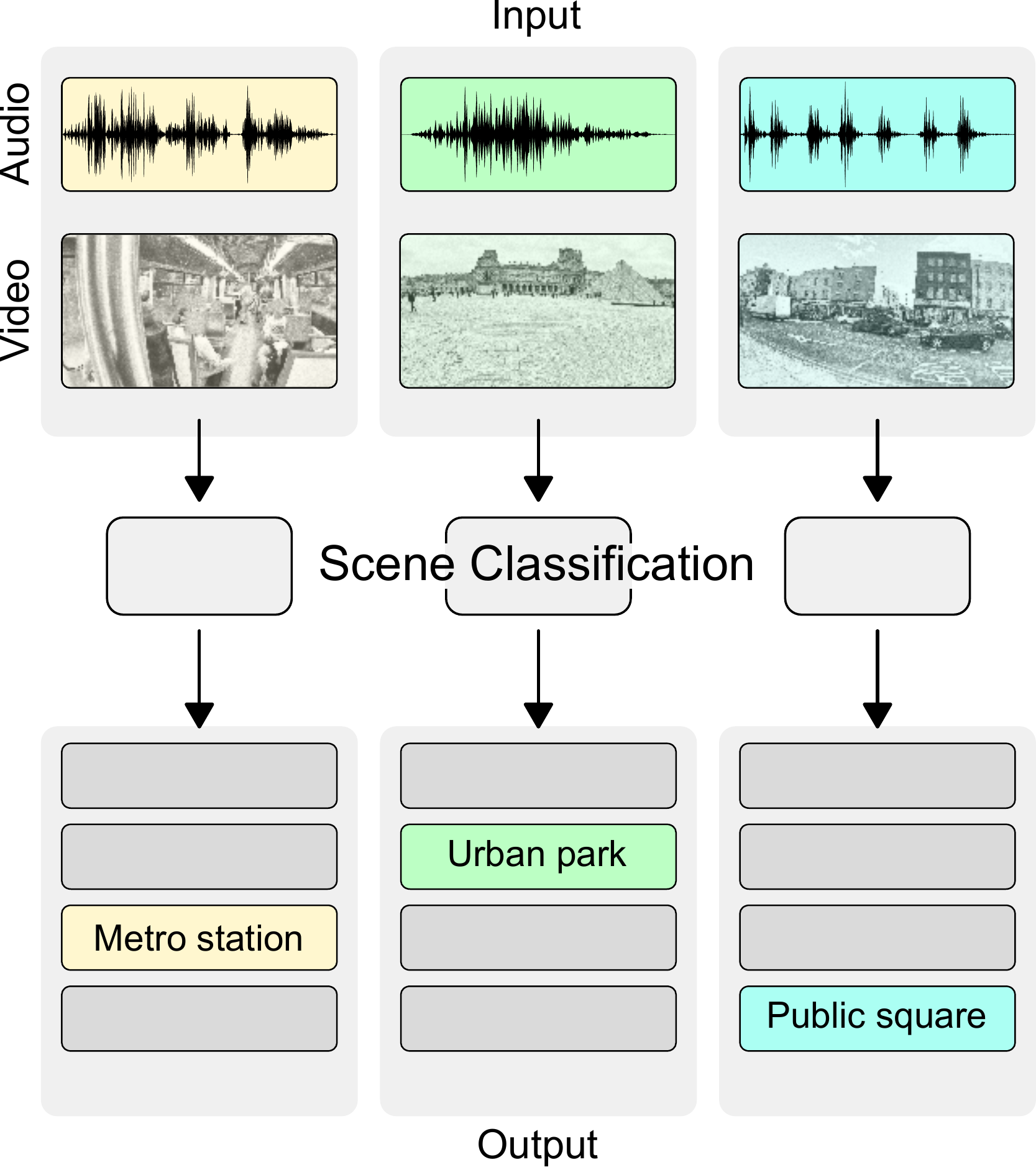}
    \caption{Overview of audio-visual scene classification}
    \label{fig:overview}
    \vspace{-10pt}
\end{figure}

\section{Audio-visual scene classification setup}

In DCASE 2021 Challenge, an audio-visual scene  classification task is introduced for the first time. The problem is illustrated in Fig.\ref{fig:overview}. The input to the system is both acoustic and visual signals. 
Approaches are allowed to use a single modality or both in order to take a decision as to what scene the audio and parallel video recordings were captured. 

\subsection{Dataset}

The dataset for this task is \textbf{TAU Urban Audio-Visual Scenes 2021}. The dataset was recorded during 2018-2019 and consists of ten scene classes: airport, shopping mall (indoor), metro station (underground), pedestrian street, public square, street (traffic), traveling by tram, bus and metro (underground), and urban park, from 12 European cities: Amsterdam, Barcelona, Helsinki, Lisbon, London, Lyon, Madrid, Milan, Prague, Paris, Stockholm, and Vienna. The audio content of the dataset is a subset of TAU Urban Acoustic Scenes 2020, in which data was recorded simultaneously with four different devices \cite{Mesaros2018_DCASE}.

The video content of the dataset was recorded using a GoPro Hero5 Session; the corresponding time-synchronized audio data was recorded using a Soundman OKM II Klassik/studio A3 electret binaural in-ear microphone and a Zoom F8 audio recorder with 48~kHz sampling rate and 24 bit resolution. The camera was mounted at chest level on the strap of the backpack, therefore the captured audio and video have a consistent relationship between moving objects and sound sources. Faces and licence plates in the video were blurred during the data postprocessing stage, to meet the requirements of the General Data Protection Regulation law by the European Union.

The development dataset contains 34 hours of data, provided in files with a length of 10 seconds. Complete statistics of the dataset content can be found in \cite{Wang2021_ICASSP}. 
The evaluation set contains 20 hours of data from 12 cities (2 cities unseen in the development set), in files with a length of 1 second.

\subsection{Performance Evaluation}
Evaluation of systems is performed using two metrics: multiclass cross-entropy (log-loss) and accuracy. Both metrics are calculated as average of the class-wise performance (macro-average), and ranking of submissions is performed using the log-loss. Accuracy is provided for comparison with the ASC evaluations from the challenge previous editions.

\subsection{Baseline system and results}

The baseline system is based on OpenL3 \cite{DBLP:journals/corr/ArandjelovicZ17} and uses both audio and video branches in the decision. The audio and video embeddings are extracted according to the original OpenL3 architecture, then each branch is trained separately for scene classification based on a single modality. The trained audio and video sub-networks are then connected using two fully-connected feed-forward layers of size 128 and 10. 

Audio embeddings are calculated with a window length of 1~s and a hop length of 0.1~s, 256 mel filters, using the "environment" content type, resulting in an audio embedding vector of length 512. Video embeddings are extracted using the same variables as the audio embedding, excluding the hop length, resulting in a video embedding  vector  of length 512. Embeddings are further preprocessed  using  z-score  normalization for bringing them to zero mean and unit variance. 
For training the joint network, Adam optimizer \cite{kingma2014adam} is used with a learning rate set to 0.0001 and weight decay of 0.0001. Cross-entropy loss is used as the loss function. The models with best validation loss are retained. More details on the system are presented in \cite{Wang2021_ICASSP}.

The baseline system results are presented in Table \ref{tab:baseline_results}. In the test stage, the system predicts an output for each 1~s segment of the data. The results from Table \ref{tab:baseline_results} are different than the ones presented in \cite{Wang2021_ICASSP}, because in the latter the evaluation is done for the 10~s clip. In that case, the final decision for a clip is based on the maximum probability over 10 classes after summing up the probabilities that the system outputs for the 1~s segments belonging to the same audio or video clip. In DCASE challenge, the evaluation data contains clips with a length of 1~s, therefore the baseline system is evaluated on segments of length 1~s also in development. 

The results in Table \ref{tab:baseline_results} show that the easiest to recognize was the "street with traffic" class, having the lowest log-loss of all classes at 0.296, and an accuracy of 89.6\%. At the other extreme is the "airport" class, with a log-loss of 0.963 and accuracy 66.8\%, and an average performance of 0.658 log-loss, with 77.0\% accuracy.

\begin{table}[]
    \small
    \centering
    \begin{tabular}{l|c}
    \toprule
     & Baseline (audio-visual)  \\
    \midrule
    Scene class &  Log loss Accuracy   \\
    \midrule
    Airport & 0.963  \hspace{3mm} 66.8\% \\
    Bus & 0.396 \hspace{3mm} 85.9\%\\
    Metro & 0.541 \hspace{3mm} 80.4\%\\
    Metro station & 0.565 \hspace{3mm} 80.8\%\\
    Park & 0.710 \hspace{3mm} 77.2\%\\
    Public square & 0.732 \hspace{3mm} 71.1\%\\
    Shopping mall & 0.839 \hspace{3mm} 72.6\%\\
    Street pedestrian & 0.877 \hspace{3mm} 72.7\%\\
    Street traffic & 0.296 \hspace{3mm} 89.6\%\\
    Tram & 0.659 \hspace{3mm} 73.1\%\\
    \midrule
    \textbf{Average} & \textbf{0.658} \hspace{3mm} 77.0\%\\
    
    \bottomrule
    \end{tabular}
    \caption{Baseline system performance on the development dataset}
    \label{tab:baseline_results}
    \vspace{-10pt}
\end{table}

\section{Challenge results}
\label{sec:results}

There are altogether 13 teams that participated to this task with one to four submission entries from each team, summing up to 43 entries. Of these, systems of 8 teams outperformed the  baseline system. The top system, Zhang\_IOA\_3 \cite{Wang2021b}, achieved a log loss of 0.195 and accuracy of 93.8\%. Among all submissions, 15 systems achieved an accuracy higher than 90\% and a log loss under 0.34. There are 11 systems which  use only the audio modality, three that use only video, and 27 multimodal systems. The best performing audio-only system, Naranjo-Alcazar\_UV\_3 \cite{Naranjo-Alcazar2021_t1b}, is  ranked 32nd with 1.006 logloss and 66.8\% accuracy. The best performing video-only system, Okazaki\_LDSLVision\_1 \cite{Okazaki2021}, is ranked 12th with 0.312 log loss and 91.6\% accuracy, while the top 8 systems belong to 2 teams, and are all multimodal.

\section{Analysis of submissions}

\begin{table*}
    \centering
    \begin{tabular}{c|l|cc|cc}
    \toprule
         Rank & Team & Logloss & Accuracy  (95\% CI)&  Fusion Methods & Model Complexity \\
         \midrule
         1 & Zhang\_IOA\_3 &0.195 &	93.8\% (93.6 - 93.9) & early fusion&110M \\
         5 & Du\_USTC\_4 & 0.221 &	93.2\% (93.0 - 93.4)  & early fusion & 373M\\
         9 & Okazaki\_LDSLVision\_4 &0.257 & 93.5\% (93.3 - 93.7)  & audio-visual& 636M \\
         10 & Yang\_THU\_3 &0.279 &	92.1\% (91.9 - 92.3)  & early fusion & 121M\\
         16 & Hou\_UGent\_4 & 0.416 &	85.6\% (85.3 - 85.8) & late fusion & 28M\\
        24 & DCASE2021 baseline & 0.662 & 77.1\% (76.8 - 77.5)  & early fusion & 711k \\
         \bottomrule
    \end{tabular}
    \caption{Performance and general characteristics of top 5 teams (best system of each team). All these systems use both audio and video.}
    \label{tab:top5 teams}
\end{table*}

A general analysis of the submitted systems shows that the most popular approach is usage of both modalities, with multimodal approaches being used by 26 of the 43 systems. Log-mel energies are the most widely used acoustic features among the submissions. Data augmentation techniques, including mixup, SpecAugment, color jitter, and frequency masking, are applied in almost every submitted system. The usage of large pretrained models  such  as  ResNet, VGG, EfficientNet trained on ImageNet or Places365 and fine-tuned on the challenge dataset is employed in most systems to extract the video embeddings. The combination of information from the audio and video modalities is implemented as both early and late fusion. The main characteristics and performance on the evaluation set of the systems submitted by the top 5 teams are presented in Table \ref{tab:top5 teams}.

\subsection{Characteristics of top systems}

The top ranked system \cite{Wang2021b} adopts multimodality to solve the task. In the audio branch, authors employed 1D deep convolutional neural network and investigated three different acoustic features: mel filter bank, scalogram extracted by wavelets, and bark filter bank, calculated from the average and difference channels, instead of left and right channels. In the video branch, authors studied four different pretrained models: ResNet-50, EfficientNet-b5, EfficientNetV2-small, and swin transformer. Authors use the pretrained model trained on ImageNet, and fine-tune it first on Places365, then on TAU Urban Audio-Visual Scenes 2021 dataset. RandomResizedCrop, RandomHorizontalFlip, and Mixup data augmentation techniques are also applied. This approach takes the top 4 ranks, with the best system being based on the combination of EfficientNet-b5 and log-mel acoustic features, a hybrid fusion comprised of model-level and decision-level fusion.

The team ranked on second place \cite{Wang2021} used an audio-visual system  and explored various pretrained models for both audio and video domain. The systems also include data augmentation through SpecAugment, channel confusion, and pitch shifting. Specifically, for the audio embedding, authors investugated use of the pretrained VGGish and PANN networks, both trained on AudioSet, with transfer learning applied to solve the AVSC task. Authors propose use of FCNN and ResNet to extract high-level audio features, to better leverage the acoustic presentations in these models. For the video embeddings, authors adopt the pretrained model trained on ImageNet and Places365. Authors also propose to use embeddings extracted from an audio-visual segment model (AVSM), to represent a scene as a temporal sequence of fundamental units by using acoustic and visual features simultaneously. The AVSM sequence is translated into embedding through a text categorization method, and authors call this a text embedding. The combination of audio, video, and text embeddings significantly improves their system's performance compared to audio-video only.

The team ranked third \cite{Okazaki2021} also used audio, video and text for solving the given task. Authors use log-mel spectrogram, frame-wise image features, and text-guided frame-wise features. For audio, the popular pretrained CNN model trained on AudioSet is used, with log-mel spectrogram;  for video, authors select three backbones ResNeSt, RegNet, and HRNet; finally, for the text modality, authors use CLIP image encoders trained on image and text caption pairs using contrastive learning, to obtain text-guided frame-wise image features. The three domain-specific models were ensembled using the class-wise confidences of the separate outputs, and postprocessed using the confidence replacement approach of thresholding the log-loss. In this way, the system can avoid the large log-loss value corresponding to a very small confidence. Authors show that this approach has significantly improved the log-loss results.

\subsection{Systems combinations}

Confidence intervals for the accuracy of the top systems presented in Table \ref{tab:top5 teams} are mostly not overlapping (small overlap between ranks 6 and 9). Logloss confidence intervals are $\pm 0.02$ for all systems. Because the systems are significantly different, we investigate some systems combinations. We first calculate the performance when combining the outputs of the top three systems with a majority vote rule. The obtained accuracy is 94.9\%, with a 95\%CI of $\pm 0.2$. Even though modest, this increase is statistically significant, showing that the systems behave differently for some of the test examples.

Looking at the same systems as a best case scenario, we calculate accuracy by considering a correct item if \textit{at least one} of the systems has classified it correctly. In this case, we  obtain an accuracy of 97.5\% with a 95\% CI of  $\pm 0.1$, showing that the vast majority of the test clips are correctly classified by at least one of the three considered systems, and that if the right rules for fusion can be found, performance can be brought very close to 100\%.

\begin{figure*}
    \centering
    \includegraphics[scale = 0.5]{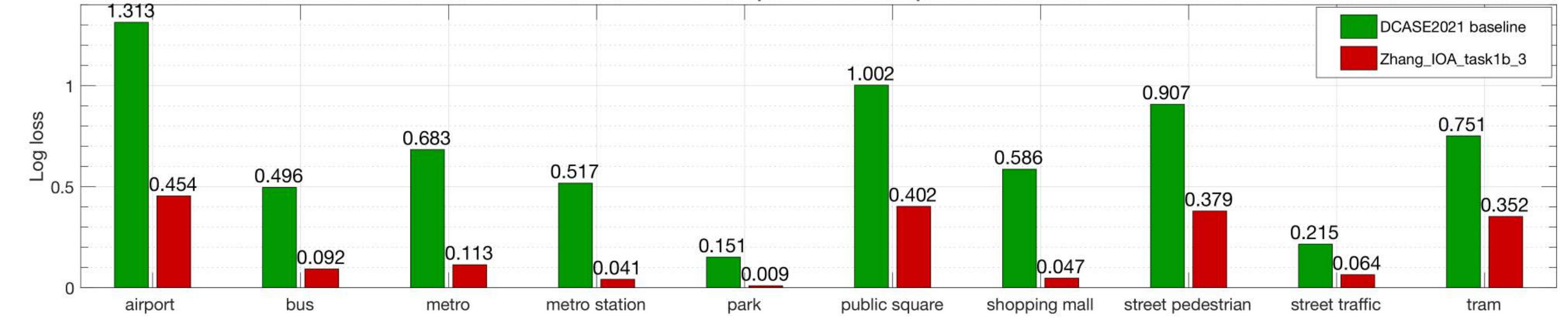}
    \vspace{-10pt}
    \caption{Class-wise performance comparison between the top 1 system and the baseline system on the evaluation set. }
    \label{fig:bar_plot}
\end{figure*}

\subsection{General trends}

An analysis over the individual modalities among the submissions reveals that video-based methods have advantages over audio-based ones. The best audio-only model achieves a logloss of 1.006 and accuracy of 66.8\%, while the best video-only model has a much lower logloss of 0.132 and much higher accuracy of 91.6\%. This is due to several reasons. Firstly, image classification domain has a relatively longer history than the audio scene classification, which allowed the development of mature and large pretrained models with millions of parameters, for example, CNNs, ResNet, VGG, EfficientNet and so forth. Secondly, the large-scale image datasets and the variety of the available image data, such as ImageNet, COCO and so forth, help making the model more robust. Thirdly, image domain has attracted much attention throughout the years, including large numbers of participants in various challenges, such as Kaggle challenges, therefore promoting the rapid development of this field.

Even though the audio-only models achieve lower performance than video-only ones, the best performance was obtained by systems which combined the two modalities. This validates our initial idea that joint modeling of audio and visual modalities can bring significant performance gain compared to state-of-the-art uni-modal systems. 

\begin{figure}
    \centering
    \includegraphics[scale = 0.55]{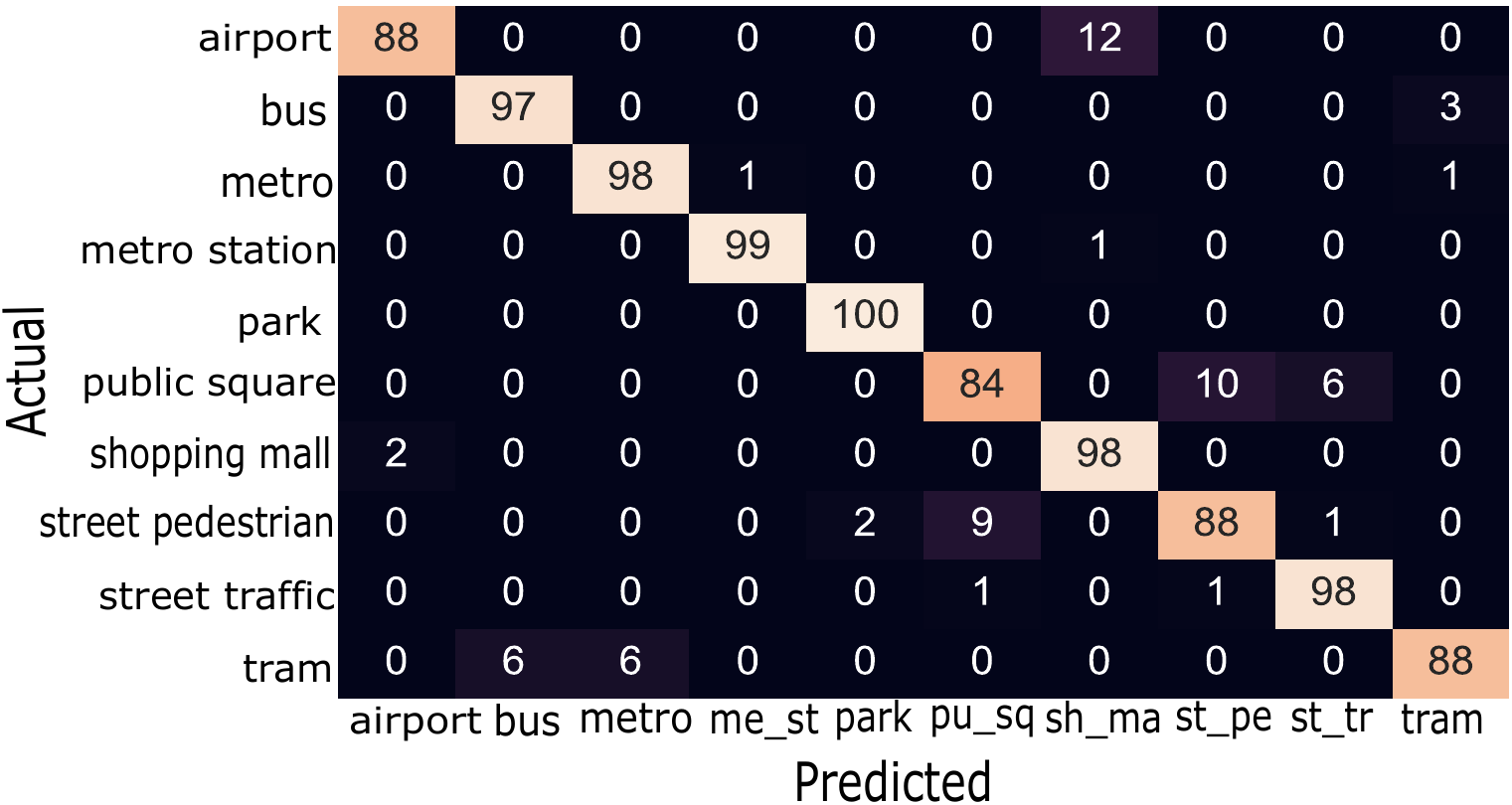}
    \caption{Confusion matrix of the top-performing system \cite{Wang2021b}. }
    \label{fig:cm}
    \vspace{-10pt}
\end{figure}

An analysis of the machine learning characteristics of the submitted systems reveals that there is a direct relationship between the performance and the model complexity, that is, in general, the top-performing systems tend to have more complex models with larger numbers of trainable parameters. 
Indeed, Spearman’s rank correlation coefficient \cite{spearman1904proof} between the model complexity and system rank is 0.75, indicating that they are highly correlated.
Considering both complexity and the performance, the baseline system is a balanced choice with a satisfactory performance.

The choice of evaluation metrics does not affect the ranking drastically. We found that the top team stays the same position, the system Okazaki\_LDSLVision\_4 would jump to the second instead of the third spot, Hou\_UGent\_4 would drop to the seventh instead of the fifth, and  Wang\_BIT\_1 would jump to the tenth from the thirteenth. The Spearman's rank correlation between accuracy and logloss indicates a very strong correlation, at 0.93.

In general, no significant changes have been found in terms of the system performance between the development dataset and the evaluation dataset, which shows that the dataset is well balanced and the systems have consistent behavior and good generalization properties. Most of the system performance shows only a very slight drop in performance on the evaluation dataset, which is explained by the data from two cities unseen in training.

The confusion matrix of the top system is shown in Fig.\ref{fig:cm}. In general, the top system performs very good in all classes; the minimum class performance is 84\%, and the highest is 100\%. In particular, the system achieves the best performance in "park"(100) and excellent in "metro station"(99), "metro"(98), "shopping mall"(98), "street traffic"(98), "bus"(97). 
We observe that "airport" class is mostly misclassified as "shopping mall"(12);  "public square" is  often misclassified as "street pedestrian"(10) and "street traffic"(6); and "tram" is mostly misclassified as "bus" and "metro". This behavior is rather intuitive, since inside the airport there are many shops which may resemble "shopping mall", and inside the "tram" there are mostly seats and people which may also resemble "bus" or "metro". 

A bar plot comparison of the class-wise performance on the evaluation set between the baseline and the top system is shown in Fig.\ref{fig:bar_plot}. 
It can be seen that the top system has significantly higher performance in all classes, especially "airport" (logloss 0.859 smaller) and
"public square" (logloss 0.6 smaller). Some similarities between the baseline system and the top system can be observed in the bar plot, with  "park" being the easiest to classify among all classes for both systems, and "airport" being the most difficult one.

\section{Conclusions and future work}
\label{sec:conclusions}
Audio-visual scene classification task was introduced in the DCASE2021 challenge for the first time, and had a high number participants and submissions. More than half of the submissions outperformed the baseline system. Multimodal approaches were widely applied among the submissions, and also achieved the best performance compared to uni-modal methods. The choice of models used by the top systems reveals that large and well-trained pretrained models are important for this task, while data augmentation and fine-tuning techniques help making the system more robust.

\bibliographystyle{IEEEtran}
\bibliography{avsc_dcase2021}

\end{sloppy}
\end{document}